\documentclass[prl,aps,showpacs,twocolumn]{revtex4}
\usepackage{epsfig,amssymb,amsfonts}

\def\opone{\leavevmode\hbox{\small1\kern-3.8pt\normalsize1}}

\begin{document}

\title{Wave Packet for Massless Fermions and its Implication to the Superluminal Velocity Statistics of Neutrino}
\author{Kelin Wang}
\affiliation{Department of Modern Physics, University of Science and Technology of China, Hefei 230026, China}
\author{ Zexian Cao}
\affiliation{Institute of Physics, Chinese Academy of Sciences, Beijing 100190, China}

\begin{abstract}
Non-dispersive wave packet for massless fermions is formulated on the basis of squeezed coherent states that are put in a form of common eigenfunction for the Hamiltonian and the helicity operator, starting from the Dirac equation. The wave packet thus constructed is demonstrated to propagate at a constant velocity as that of light. This explicit expression of wave packet for the massless fermions can facilitate theoretical analysis of problems where a wave packet is of formal significance. Furthermore, extensive wave packet may result in a superluminal velocity statistics if determined from the time-of-flight measurement, as recently done on muon neutrinos, when a threshold particle flux or energy transfer, which is eventually referred to the propagation of wave packet, to invoke a detection event is assumed.
\end{abstract}

\pacs{03. 65. Pm; 05. 30. Fk; 06. 30. Gv; 14. 60. Lm}

\maketitle

In fundamental quantum theory, a free particle is represented by a plane wave $e^{i(kx-\omega t)}$ which is the eigenfunction of the momentum operator. However, the plane wave cannot be normalized, thus it does not fit our picture of particles which at least should make their presence only in a limited region of microscopic scale \cite{ldl}. Packet waves have been formulated for light coming in indivisible unit of photons, a kind of massless boson, and for massive fermions such as electron, based on a dispersion relation derived from the electromagnetic wave equation for the former and from the Schr$\ddot{o}$dinger equation or Dirac equation for the latter, respectively. Remarkably, however, there is no mathematical formulation of wave packet for massless fermions, to the best knowledge of the authors. With the absence of massless fermions, the picture of particles as a wave packet is obviously incomplete.

With the introduction of wave packet, various particle behaviors are often discussed as a matter of fact in terms of this concept. Yet in many circumstances, for instance when talking about the collision or detection by a detector, the particles are simply thought of as a point object. The handling of particles as a point object is a valid convenience when the size of the particles is much smaller than the scale of the experimental setup, or to be nearer to the reality,less than the size of atoms or atomic clusters that constitute the responsive agent of a detector. This is not a challenging requirement for electrons or protons since these massive fermions have a very short Compton wavelength and they just behave as distinct identities though formally they can also be represented as a wave packet. For such massive fermions, the concepts of point particle and wave packet do not cause much confusion, and the output of measurement for one single particle striking a macroscopic detector should give a sharp value within the precision of the detection method. In the case of massless bosons, things become complicated. Though we know that light comes in indivisible unit of photons, our description of relationship between light and photon is incomplete since, for example, the light can be polarized but we don't know how each photon can keep a record of the light's polarity. This would take us away from the traditional ideal of elementary particles, and more significance may be attributed to the concept of wave pack. Put this fact aside, we notice that the photons now can be brought into light pulses of enormously differing durations. It is reasonably conceivable that the wave packets describing light can in principle assume a macroscopic extension, noting that a temporal duration of 1.0 ns corresponds to a spatial extension of 30 cm for a light pulse subject to none of the decoherence mechanisms.

A quite natural question may be posed about massless fermions: how could they be represented as a packet wave, and what the extensive wave packet may imply for the investigation of behaviors of such particles? This may be a question of some  relevance, say, with regard to the recent report of superluminal velocity measurement on muon neutrinos \cite{tad}. Neutrino is a massless fermion in the standard model. Some experimental observations, e.g., neutrino oscillation, point to the probability that neutrinos may have mass. It is believed that the upper limit for heavier neutrinos is at the order of magnitude of 2 eV, and the lightest neutrino could in principle be massless \cite{chw}. We shall always presuppose $m_{\nu}=0$ without distinguish the flavors of the neutrinos in the current work. Now, a massless neutrino, following relativistic quantum mechanics, should propagate at the velocity of light in vacuum, c, and the detection of its velocity, if the neutrino behaves as a point particle, is expected to give a sharp value of c within the precision of the method.  However, Adam and coworkers recently showed that the velocity of muon neutrino with the average energy of 17 GeV, by time-of-flight measurement, displays surprisingly a skewed distribution biased at the $>c$ side. This has been even interpreted as the proof of superluminal existence. A detailed inspection of the data for the experiment performed at the average of 28.1 GeV further indicates that the data taking only events at energy $>$20 GeV give to a more severe superluminal velocity statistics \cite{tad}. Could it be possible that the massless neutrinos propagate as a train of wave at a group velocity of c, yet the behavior of the wave packet could lead to an energy-dependent velocity statistics deviating from c?

In this article, we report the formulation of wave packet, starting from the Dirac equation, for massless fermions on the basis of the squeezed coherent states \cite{zhw}, and the wave packet thus constructed is non-dispersive and propagates at the velocity of light. We see that by assuming a threshold particle flux or energy transfer, which to the end is referred to the probability distribution described by the wave packet, is desired to invoke a detection event for the particle, the measured velocity statistics for neutrinos could find a reasonable explanation without breaking relativity.

For a massless fermion, the Hamiltonian is given in the form
\begin{equation}
H=c\overrightarrow{\sigma}\cdot\overrightarrow{P}.
\end{equation}
Where, c, the velocity of light, is a parameter that connects space and time. This doesn't necessarily imply that the particle has a definite velocity as that of light, bearing in mind the Dirac equation for the massive electron.

For simplicity, without loss of generality, we consider the situation that the particle propagates along x-direction, then the Hamitonian (1) is rewritten as
\begin{equation}
H=c\sigma_{x}\hat{p}_{x}.
\end{equation}
Now, let's represent the position operator $\hat{x}$  and momentum $\hat{p}$ in terms of the creation and annihilation operators
\begin{equation}
\hat{x}=\frac{i\Delta}{\sqrt{2}}(b-b^{\dagger}),
\hat{p}=\frac{1}{\sqrt{2}\Delta}(b^{\dagger}+b).
\end{equation}
With $\Delta$ being a parameter in the dimension of length, then the Hamitonian in Eq.(2) turns into
\begin{equation}
H=A\sigma_{x}(b+b^{\dagger}),
\end{equation}
Where $A=\frac{c}{\sqrt{2}\Delta}$.

It is obvious that the helicity, $\eta=\overrightarrow{\sigma}\cdot\overrightarrow{P}/\texttt{p}$, i.e., the projection of the spin of the particle onto the direction of momentum, is a conserved quantity since it commutates with the Hamiltonian in Eq.(2).  The eigenstates for the helicity operator are given as follows
\begin{eqnarray}
|+\rangle=\left(\begin{array}{c}
                    (e^{-\frac{1}{2}b^{\dagger}b^{\dagger}+b^{\dagger}})|0\rangle \\
                    (e^{-\frac{1}{2}b^{\dagger}b^{\dagger}+b^{\dagger}})|0\rangle
                  \end{array}
\right),\nonumber\\
|-\rangle=\left(\begin{array}{c}
                    (e^{-\frac{1}{2}b^{\dagger}b^{\dagger}-b^{\dagger}})|0\rangle \\
                    (e^{-\frac{1}{2}b^{\dagger}b^{\dagger}-b^{\dagger}})|0\rangle
                  \end{array}
\right).
\end{eqnarray}
which correspond to the eigenvalue +1 and -1, respectively.
The eigenstates in Eq.(5) are, however, not normalizable since the integral arising from the term $e^{\beta b^{\dagger}b^{\dagger}}|0\rangle$ is unlimited for $|\beta|\geq\frac{1}{2}$. Consequently, such eigenstates of a unique helicity is not proper for the physical representation of a massless fermon though it was claimed that the state of a massless neutrino could never be a superposition of two different helicities and the experiment gave a direct indication that the helicity of neutrino is negative \cite{asg}.  In order to construct a normalizable physical state in the form of wave packet for a massless fermion, the mixing of helicity seems unavoidable. Let's take the wave packet given below as the initial state of the massless fermion
\begin{eqnarray}
|t&=&0\rangle=\int e^{-\gamma(\alpha-\alpha_{0})^{2}}\left(
                                       \begin{array}{c}
                                         e^{-\frac{1}{2}b^{\dagger}b^{\dagger}+ab^{\dagger}}|0\rangle\\
                                         e^{-\frac{1}{2}b^{\dagger}b^{\dagger}+ab^{\dagger}}|0\rangle \\
                                       \end{array}
                                     \right)\texttt{d}\alpha,
\end{eqnarray}
where the weight factor in the integral is Gaussian. It can be shown that such an initial state is normalizable, as can be proved that
 $\langle t=0|t=0\rangle=\frac{4\pi}{\sqrt{4\gamma-1}}e^{\frac{2\gamma}{4\gamma-1}\alpha_{0}^{2}}$, in deriving this we have used the enclosure relation for the coherent state $|\Phi\rangle=e^{{\phi}b^{\dagger}}|0\rangle$,  i.e., $\int\frac{\texttt{d}(\texttt{Re}\Phi)\texttt{d}(\texttt{Im}\Phi)}{\pi}e^{-(\texttt{Re}\Phi^{2}+\texttt{Im}\Phi^{2})}|\Phi\rangle\langle\Phi|=1$, and the integrals $\int^{\infty}_{-\infty}e^{-\beta x^{2}}\sin(bx)\texttt{d}x=0$ and $\int^{\infty}_{-\infty}e^{-\beta x^{2}}\cos(bx)\texttt{d}x=\sqrt{\frac{\pi}{\beta}}e^{\frac{b^{2}}{4\beta}}$.
 
The temporal evolution of the state represented by the wave packet in Eq.(6) can be derived in a straightforward way. At any moment $t$, it has
\begin{equation}
|t\rangle=\int e^{-\gamma(\alpha-\alpha_{0})^{2}}e^{-iE(\alpha)t}\left(
                                                                   \begin{array}{c}
                                                                     e^{-\frac{1}{2}b^{\dagger}b^{\dagger}+ab^{\dagger}}|0\rangle \\
                                                                     e^{-\frac{1}{2}b^{\dagger}b^{\dagger}+ab^{\dagger}}|0\rangle  \\
                                                                   \end{array}
                                                                 \right)\texttt{d}\alpha.
\end{equation}

Thus the expectation value for the position operator at any moment $t$ is
\begin{equation}
\overline{x}(t)=\frac{\langle t|\hat{x}|t\rangle}{\langle t=0|t=0\rangle}=\frac{\langle t|\frac{i\Delta}{\sqrt{2}}(b-b^{\dagger})|t\rangle}{\langle t=0|t=0\rangle},
\end{equation}
which leads to the result $\overline{x}(t)=\sqrt{2}\Delta\cdot At$, implying that the velocity of the packet wave is a constant of \emph{c}, as $v=\frac{\overline{x}(t)}{t}=\sqrt{2}\Delta\cdot A=c$.

Now let's calculate the width of the wave packet at any moment $t$, $\Gamma=(\langle\hat{x}(t)^{2}\rangle-\langle\hat{x}(t)\rangle^{2})^{1/2}$. Note that $\langle t|\hat{x}^{2}|t\rangle=-\frac{\Delta^{2}}{2}\langle t|(b-b^{\dagger})^{2}|t\rangle$, it thus has $\langle\hat{x}(t)^{2}\rangle=\frac{\langle t|x^{2}|t\rangle}{\langle t=0|t=0\rangle}=2\Delta^{2}(A^{2}t^{2}+\frac{1}{\gamma})-\frac{\Delta^{2}}{2}$, consequently
\begin{equation}
\Gamma(t)=[2\Delta^{2}(A^{2}t^{2}+\frac{1}{\gamma})-\frac{\Delta^{2}}{2}-(\sqrt{2}\Delta At)^{2}]^{1/2}=\Delta(\frac{2}{\gamma}-\frac{1}{2})^{1/2}.
\end{equation}
Based on the arguments above one concludes that the wave packet given in Eq.(6) is non-dispersive, and propagates at a velocity of \emph{c}, therefore it can be used to represent a massless fermion or a beam of massless fermions.It is noteworthy to mention that $4\geq\gamma\geq1/4$ as required by the normalization condition and the positiveness of the wave packet width.

From Eq.(9) we see that the extension of this non-dispersive wave packet is proportional to the free parameters $\Delta$, consequently it can in principle be made sufficiently extensive. This is analogous to the case of light for which the wave packet can be of an enormously variable lengths, since the Compton wavelength of photon is infinitely long and the photons are almost decoupled from each other. For massless particles such as photons and neutrinos, their behaviors are expected to display more undulatory characteristics. In contrast, massive particles such as electrons and neutrons have a limited Compton wavelength, their wave packet can have only a limited extension-in general cases they are just treated as a distinct particle instead of a train of wave.

Now it is  probably appropriate to discuss the statistics of velocity measurement for the massless fermions, e.g., for neutrino. When the neutrinos are represented by the wave packet formulated in Eq.(6), does that imply the possibility of measuring a superluminal velocity under certain circumstances? Or, what kind of an energy-dependent velocity measurement can be anticipated for neutrinos when taken as a non-dispersive wave packet with a constant propagation velocity of \emph{c}?

The difference of the measured velocities for muon neutrinos relative to that of light has been given in a statistical notion, and the original data of velocity scatter randomly around the velocity of light. For the muon neutrinos with an average energy of 17 GeV, Adam and coworkers, based on the measurement of time-of-flight through a base line of 730 km long, came to the conclusion that $(\texttt{v}-c)/c=(2.48\pm0.28(\texttt{stat}.)\pm0.30(\texttt{sys}.))\times10^{-5}$\cite{tad}. A measurement performed for an average neutrino energy of 28.1 GeV shows only negligible difference from the result above, but the statistics taking events of energy higher than 20 GeV shows a significantly larger velocity than that for energy below 20 GeV, the early neutrino arrival time $\delta t$ is $\sim$67.1 ns in the former case while only $\sim$53.1ns in the latter case \cite{tad}. Furthermore, it is worth noting that the Minos experiment with neutrino energy peaking at $\sim$3.0 GeV with a tail extending to above 100 GeV results in $(\texttt{v}-c)/c=5.1\pm2.9\times10^{-5}$ \cite{pad}, but a stringent limit of $(\texttt{v}-c)/c<2\times10^{-9}$ (given in absolute value!) was set by the observation of (anti) neutrinos emitted by the SN1987A supernova in the 100 MeV range \cite{khi,rmb}. Roughly speaking, the measured neutrino velocity seems to suggest that following a larger neutrino energy there may be an increased probability of measuring a superluminal velocity.

The wave packet formulated in Eq.(6) for massless fermions is strictly non-dispersive, and it propagates at a constant velocity of light. This concludes from the solution of the equation of motion for the particles, which is consistent with both special relativity and quantum theory. If the velocity of such a wave packet is determined by measuring the time of flight, then so long as time tagging is performed with respect to precisely the same position on the wave packet, say the center of the wave packet but not necessarily, it will result in the same value, \emph{c}, even though it has a finite extension. For the muon neutrinos in Ref.\cite{tad}, it requires a time tagging referring to exactly the same position of the wave packet at the source and at the detector, a quite challenging task because the time tagging was performed via different processes. The finite extension of the wave packet may cause a peculiar systematic deviation of measured neutrino velocity from the velocity of light, following the Copenhagen interpretation of the wave function.

The event that a particle initiates as being registered, whether on the surface of a solid detector directly or via an intermediate nuclear reaction, is essentially quantum mechanical, i.e., in a statistical sense. Suppose that it needs a threshold particle flux, or energy transfer, to invoke a detection event, e.g., to generate taus to be measured to indicate the arrival of the neutrinos \cite{tad}, then this implies that a threshold portion of the wave packet has to enter the detection region or the device \cite{yah,jba}. As we mentioned above, the wave packet for neutrinos can have a macroscopic extension as a pulse of light, the superluminal velocity statistics for neutrino can find a natural and straightforward explanation under the threshold probability assumption. In case a larger threshold portion of the wave packet, implying a larger particle flux or energy transfer, is required to invoke a detection event that the time tagging at the detector might be retarded relative to the time tagging at the source, the velocity distribution will be biased by $<c$  results; and if a smaller threshold probability of the wave packet is required that  the time-of flight measurement may lead to an early arrival time, then the velocity distribution may be biased by $>c$  results (Figure 1). Clearly, for the beam of neutrinos of greater energy, the velocity distribution may contain more superluminal data, farther from the value of \emph{c}.

In summary, non-dispersive wave packet with a constant propagation velocity of \emph{c} has been formulated for massless fermions starting from the Dirac equation. Such a wave packet is constructed from the superposition of common eigenstates of the Hamiltonian and the helicity operator, which are represented on the basis of the squeezed-coherent states. Based on the assumption that a threshold portion of the wave packet is demanded to invoke a detection event, the wave packet, macroscopically extensive as it may be in principle as a beam of light, can provide a quite reasonable explanation to the superluminal velocity statistics recently reported for neutrino, which is anticipated to occur more frequently and farther away from the velocity of light at higher particle energies. The result can serve as a basis for the discussion of other problems concerning massless fermions where the formulation of the wave packet may be of formal significance.

Acknowledgment. One of the author (Cao) thanks the financial supports from the national natural science foundation of China grant no. 10974227, the ministry of science and technology grant no. 2009CB930800 , and the Chinese academy of sciences.

\begin{figure}
\begin{center}
\epsfig{figure=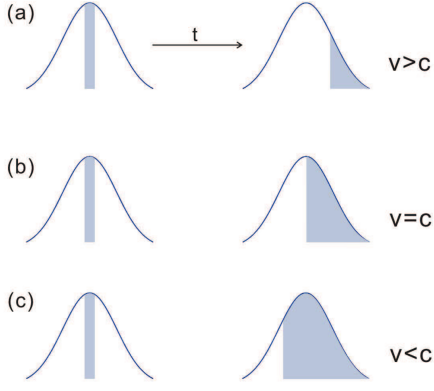,width=0.35\textwidth}
\end{center}
\caption{Velocity measurement of a non-dispersive wave packet for massless fermion with a constant propagation velocity of \emph{c}. In case that the portion of the incident wave packet demanded to invoke a detection event is (a) ahead of, (b) coincident with, or (c) lagging behind the point registered as $t=0$ , then the velocity calculated from the time of flight will be $\texttt{v}>c$, $\texttt{v}=c$ or $\texttt{v}<c$, respectively.}
\end{figure}

\section{Appendix}

1. The module square of the initial state $|t=0\rangle$

For
\begin{eqnarray}
|t&=&0\rangle=\int e^{-\gamma(\alpha-\alpha_{0})^{2}}\left(
                                       \begin{array}{c}
                                         e^{-\frac{1}{2}b^{\dagger}b^{\dagger}+ab^{\dagger}}|0\rangle\\
                                         e^{-\frac{1}{2}b^{\dagger}b^{\dagger}+ab^{\dagger}}|0\rangle \\
                                       \end{array}
                                     \right)\texttt{d}\alpha\nonumber,
\end{eqnarray}
\begin{widetext}
\begin{eqnarray}
\langle t=0|t=0\rangle&=&\int e^{-\gamma(\alpha-\alpha_{0})^{2}-\gamma(\alpha^{'2}-\alpha_{0})^{2}}\texttt{d}\alpha \texttt{d}\alpha^{'}\cdot2\langle0|e^{-\frac{1}{2}bb+\alpha b}e^{-\frac{1}{2}b^{\dagger}b^{\dagger}+\alpha^{'}b^{\dagger}}|0\rangle\nonumber\\
&=&\int e^{-\gamma(\alpha-\alpha_{0})^{2}-\gamma(\alpha^{'2}-\alpha_{0})^{2}}\texttt{d}\alpha \texttt{d}\alpha^{'}\cdot2\langle0|e^{-\frac{1}{2}b b+\alpha b}\int\frac{\texttt{d}(\texttt{Re}\Phi)\texttt{d}(\texttt{Im}\Phi)}{\pi}e^{-(\texttt{Re}\Phi^{2}+\texttt{Im}\Phi^{2})}|\Phi\rangle\langle\Phi|e^{-\frac{1}{2}b^{\dagger}b^{\dagger}+\alpha^{'}b^{\dagger}}|0\rangle\nonumber\\
&=&2\int e^{-\gamma(\alpha-\alpha_{0})^{2}-\gamma(\alpha^{'2}-\alpha_{0})^{2}}\texttt{d}\alpha \texttt{d}\alpha^{'}\int\frac{\texttt{d}(\texttt{Re}\Phi)\texttt{d}(\texttt{Im}\Phi)}{\pi}e^{-2\texttt{Re}\Phi^{2}+(\alpha+\alpha^{'})\texttt{Re}\Phi+i(\alpha-\alpha^{'})\texttt{Im}\Phi}\nonumber.
\end{eqnarray}
For brevity, replacing $\texttt{Re}\Phi$ with $\zeta$ and $\texttt{Im}\Phi$ with $\eta$,
\begin{eqnarray}
\langle t=0|t=0\rangle&=&\frac{2}{\pi}\int e^{-\gamma(\alpha-\alpha_{0})^{2}-\gamma(\alpha^{'2}-\alpha_{0})^{2}}\texttt{d}\alpha \texttt{d}\alpha^{'}\int\texttt{d}\zeta\texttt{d}\eta e^{-2\zeta^{2}+(\alpha+\alpha^{'})\zeta+i(\alpha-\alpha^{'})\eta}\nonumber\\
&=&\frac{2}{\pi}\int e^{-\gamma(\alpha-\alpha_{0})^{2}-\gamma(\alpha^{'2}-\alpha_{0})^{2}}\texttt{d}\alpha \texttt{d}\alpha^{'}\int\texttt{d}\zeta\texttt{d}\eta e^{-2\zeta^{2}+(\alpha+\alpha^{'})\zeta}[\cos(\alpha-\alpha^{'})\eta+i\sin(\alpha-\alpha^{'})\eta]\nonumber.
\end{eqnarray}
Discarding the imaginary part of the integral as it must vanish, one obtains
\begin{eqnarray}
\langle t=0|t=0\rangle&=&\frac{2}{\pi}\int e^{-\gamma(\alpha-\alpha_{0})^{2}-\gamma(\alpha^{'2}-\alpha_{0})^{2}}\texttt{d}\alpha \texttt{d}\alpha^{'}\int\texttt{d}\zeta\texttt{d}\eta e^{-2\zeta^{2}+(\alpha+\alpha^{'})\zeta}\cos(\alpha-\alpha^{'})\eta\nonumber\\
&=&\frac{2}{\pi}\int e^{-\gamma(\alpha-\alpha_{0})^{2}-\gamma(\alpha^{'2}-\alpha_{0})^{2}}\texttt{d}\alpha \texttt{d}\alpha^{'}\int\texttt{d}\zeta\texttt{d}\eta e^{-2(\zeta-\frac{\alpha+\alpha^{'}}{4})^{2}+\frac{(\alpha+\alpha^{'})^{2}}{8}}\cos(\alpha-\alpha^{'})\eta\nonumber\\
&=&\frac{2}{\pi}\int e^{-\gamma(\alpha-\alpha_{0})^{2}-\gamma(\alpha^{'2}-\alpha_{0})^{2}}\texttt{d}\alpha \texttt{d}\alpha^{'}\int\texttt{d}\zeta^{'}\texttt{d}\eta e^{-2\zeta^{'2}+\frac{(\alpha+\alpha^{'})^{2}}{8}}\cos(\alpha-\alpha^{'})\eta\nonumber\\
&=&\frac{2}{\pi}\sqrt{\frac{\pi}{2}}\int e^{-\gamma(\alpha-\alpha_{0})^{2}-\gamma(\alpha^{'2}-\alpha_{0})^{2}}\texttt{d}\alpha \texttt{d}\alpha^{'}\int\texttt{d}\eta e^{\frac{(\alpha+\alpha^{'})^{2}}{8}}\cos(\alpha-\alpha^{'})\eta\nonumber\\
&=&\sqrt{\frac{2}{\pi}}\int e^{-\gamma(\alpha^{2}-2\alpha\alpha_{0}+\alpha_{0}^{2})+\frac{\alpha^{2}}{8}+\frac{\alpha\alpha^{'}}{4}}e^{-\gamma(\alpha^{'}-\alpha_{0})^{2}+\frac{\alpha^{'2}}{8}}\cos(\alpha-\alpha^{'})\eta\texttt{d}\alpha\texttt{d}\alpha^{'}\texttt{d}\eta\nonumber\\
&=&\sqrt{\frac{2}{\pi}}e^{-\gamma\alpha_{0}^{2}}\int e^{-(\gamma-\frac{1}{8})\alpha^{2}+\frac{8\gamma\alpha_{0}+\alpha^{'}}{4}\alpha} e^{-\gamma(\alpha^{'}-\alpha_{0})^{2}+\frac{\alpha^{'2}}{8}}\cos(\alpha-\alpha^{'})\eta\texttt{d}\alpha\texttt{d}\alpha^{'}\texttt{d}\eta\nonumber\\
&=&\sqrt{\frac{2}{\pi}}e^{-\gamma\alpha_{0}^{2}}\int e^{-(\gamma-\frac{1}{8})\alpha^{''2}}\cos(\alpha^{''}+\frac{8\gamma\alpha_{0}+\alpha^{'}}{8(\gamma-\frac{1}{8})}-\alpha^{'})\eta\texttt{d}\alpha^{''}e^{\frac{(8\gamma\alpha_{0}+\alpha^{'})^{2}}{8(8\gamma-1)}-\gamma(\alpha^{'}-\alpha_{0})^{2}+\frac{\alpha^{'2}}{8}}\texttt{d}\alpha^{'}\texttt{d}\eta\nonumber\\
&=&\sqrt{\frac{2}{\pi}}e^{-\frac{8\gamma^{2}-2\gamma}{8\gamma-1}}\alpha_{0}^{2}\int e^{-(\gamma-\frac{1}{8})\alpha^{''2}}[\cos\alpha^{''}\eta\cos\frac{8\gamma\alpha_{0}-(8\gamma-2)\alpha^{'}}{8\gamma-1}\eta\nonumber\\
&-&\sin\alpha^{''}\eta\sin\frac{8\gamma\alpha_{0}-(8\gamma-2)\alpha^{'}}{8\gamma-1}\eta]e^{\frac{1}{8\gamma-1}[-(8\gamma^{2}-2\gamma)\alpha^{'2}+16\gamma^{2}\alpha_{0}\alpha^{'}]}\texttt{d}\alpha^{''}\texttt{d}\alpha^{'}\texttt{d}\eta\nonumber\\
&=&\sqrt{\frac{2}{\pi}}e^{-\frac{8\gamma^{2}-2\gamma}{8\gamma-1}}\alpha_{0}^{2}\int\sqrt{\frac{8\pi}{8\gamma-1}}e^{-\frac{2}{8\gamma-1}\eta^{2}}\cos\frac{8\gamma\alpha_{0}-(8\gamma-2)\alpha^{'}}{8\gamma-1}\eta\nonumber\\
&\times&e^{-\frac{8\gamma^{2}-2\gamma}{8\gamma-1}(\alpha^{'}-\frac{8\gamma^{2}\alpha_{0}}{8\gamma^{2}-2\gamma})^{2}}e^{\frac{64\gamma^{4}\alpha_{0}^{2}}{(8\gamma-1)(8\gamma^{2}-2\gamma})}\texttt{d}\alpha^{'}\texttt{d}\eta\nonumber\\
&=&\sqrt{\frac{16}{8\gamma-1}}e^{\frac{4\gamma^{2}}{8\gamma^{2}-2\gamma}\alpha_{0}^{2}}\int e^{-\frac{2}{8\gamma-1}\eta^{2}}e^{-\frac{8\gamma^{2}-2\gamma}{8\gamma-1}\alpha^{'''2}}\cos[\frac{8\gamma\alpha_{0}}{8\gamma-1}-\frac{8\gamma-2}{8\gamma-1}(\alpha^{'''}+\frac{8\gamma^{2}\alpha_{0}}{8\gamma^{2}-2\gamma})]\eta\texttt{d}\alpha^{'''}\texttt{d}\eta\nonumber\\
&=&\sqrt{\frac{16}{8\gamma-1}}e^{\frac{4\gamma^{2}}{8\gamma^{2}-2\gamma}\alpha_{0}^{2}}\int e^{-\frac{2}{8\gamma-1}\eta^{2}}e^{-\frac{8\gamma^{2}-2\gamma}{8\gamma-1}\alpha^{'''2}}\cos\eta\alpha^{'''}\frac{8\gamma-2}{8\gamma-1}\texttt{d}\alpha^{'''}\texttt{d}\eta\nonumber\\
&=&\sqrt{\frac{16\pi}{8\gamma^{2}-2\gamma}}e^{\frac{4\gamma^{2}}{8\gamma^{2}-2\gamma}\alpha_{0}^{2}}\int\texttt{d}\eta e^{-\frac{1}{2\gamma}\eta^{2}}\nonumber\\
&=&\frac{4\pi}{\sqrt{4\gamma-1}}e^{\frac{2\gamma}{4\gamma-1}}\alpha_{0}^{2}.
\end{eqnarray}
\end{widetext}

2.  Motion of the wave packet

For state $|t=0\rangle$, its temporal evolution is given by
\begin{equation}
|t\rangle=\int e^{-\gamma(\alpha-\alpha_{0})^{2}}e^{-iA\alpha t}\left(
                                                                  \begin{array}{c}
                                                                    e^{-\frac{1}{2}b^{\dagger}b^{\dagger}+\alpha b^{\dagger}}|0\rangle \\
                                                                    e^{-\frac{1}{2}b^{\dagger}b^{\dagger}+\alpha b^{\dagger}}|0\rangle  \\
                                                                  \end{array}
                                                                \right)\texttt{d}\alpha.
\end{equation}

Thus it has $\overline{x}(t)\cdot\langle t=0|t=0\rangle=\langle t|\frac{i\Delta}{\sqrt{2}}(b-b^{\dagger})|t\rangle$
from which, following the same tricks in deriving (10), one obtains
\begin{equation}
\overline{x}\cdot\langle t=0|t=0\rangle=\sqrt{2}\Delta At\frac{4\pi}{\sqrt{4\gamma-1}}e^{\frac{2\gamma}{4\gamma-1}\alpha_{0}^{2}}.
\end{equation}
This implies that $\overline{x}(t)=\sqrt{2}\Delta At$, consequently the velocity of the wave packet is
\begin{equation}
v=\frac{\overline{x}(t)}{t}=\sqrt{2}\Delta A=c.
\end{equation}

3. Dispersion of the wave packet

The width of the wave packet is given by $\Gamma=[\langle t|(\hat{x}-\bar{x})^{2}|t\rangle/\langle t||t\rangle]^{1/2}$ which can be simplified to $\Gamma=[\frac{\langle t|\hat{x}^{2}|t\rangle}{\langle t=0|t=0\rangle}-\bar{x}^{2}]^{1/2}$. Since
\begin{eqnarray}
\langle t|\hat{x}^{2}|t\rangle&=&\langle t|(\frac{i\Delta}{\sqrt{2}}(b-b^{\dagger}))^{2}|t\rangle\nonumber\\
&=&-\frac{\Delta^{2}}{2}\langle t|(b^{2}+b^{\dagger 2}-2b^{\dagger}b)|t\rangle-\frac{\Delta^{2}}{2}\langle t|t\rangle.
\end{eqnarray}

Thus we just need calculate $\langle t|\hat{x}^{2}|t\rangle+\frac{\Delta^{2}}{2}\langle t|t\rangle$ which, again following the similar techniques in deriving (23), can be found to result in
\begin{equation}
\langle t|\hat{x}^{2}|t\rangle+\frac{\Delta^{2}}{2}\langle t|t\rangle=\frac{8\Delta^{2\pi}}{\sqrt{4\gamma-1}}e^{\frac{4\gamma^{2}}{8\gamma^{2}-2\gamma}\alpha_{0}^{2}}(A^{2}t^{2}+\frac{1}{\gamma}),
\end{equation}
or $\frac{\langle t|\hat{x}^{2}|t\rangle}{\langle t|t\rangle}=2\Delta^{2}(A^{2}t^{2}+\frac{1}{\gamma})-\frac{\Delta^{2}}{2}$ which further leads to $\Gamma(t)=\Delta(\frac{2}{\gamma}-\frac{1}{2})^{1/2}$, i.e., the width remains constant. The wave packet is non-dispersive.
\end{document}